# Exploring Dependence, Overreliance, and Addiction Related Behaviors Associated with Large Language Model Use Among Software Engineers

Ronnie de Souza Santos[a,*], Italo Santos[b], Matheus de Morais Leça[a], Cleyton Magalhães[c] and Mairieli Wessel[d]

[a] *University of Calgary, Calgary, Alberta, Canada*
[b] *University of Hawaii at Manoa, Honolulu, Hawaii, USA*
[c] *Universidade Federal Rural de Pernambuco (UFRPE), Recife, Brazil*
[d] *Radboud University, Nijmegen, Netherlands*

ABSTRACT

The widespread adoption of Large Language Models (LLMs) has changed how software engineers perform everyday development activities. While these systems provide substantial support for tasks such as code generation, debugging, and documentation, their increasing integration into professional workflows has also raised questions regarding developers' reliance on these tools and the emergence of dependence, overreliance, and addiction-related behaviors. This study investigates how software engineers experience the use of LLMs during professional software development, with attention to behavioral patterns associated with dependence, overreliance, and addiction-related behaviors. An exploratory survey was conducted with 119 software practitioners. The data were analyzed using descriptive statistics and qualitative thematic analysis of participants' open-ended responses. Participants primarily described functional dependence, with LLMs becoming integrated into routine software engineering activities because of the productivity and efficiency they provide. Responses also suggested patterns consistent with overreliance, particularly through prioritizing LLMs over documentation or peer consultation while continuing to verify generated outputs. Reports associated with addiction-related behaviors were less common and primarily reflected difficulty moderating use or emotional attachment to the technology rather than impaired control. The findings suggest that LLMs are becoming a habitual component of professional software engineering practice. While most reported use appears functional, the results indicate the importance of promoting appropriate reliance by supporting trust calibration, professional judgment, and verification throughout software development.

## 1. Introduction

Large Language Models (LLMs) are deep learning systems trained on large textual corpora to generate human-like responses (Hou et al., 2024; Zheng et al., 2025; Gao et al., 2025). With the increasing availability of tools such as ChatGPT and GitHub Copilot, LLMs have become embedded in software engineering workflows, supporting a wide range of activities across the development lifecycle (Ozkaya, 2023; Fan et al., 2023; Sallou et al., 2024; Santos et al., 2025). These systems are used for tasks including code generation, bug fixing, test case creation, documentation, and requirements engineering (Hou et al., 2024; Rasnayaka et al., 2024; Haque, 2025). In addition, LLMs are increasingly integrated into collaborative practices, acting as pair programming assistants or supporting code review and debugging activities during implementation (Fan et al., 2023; Ozkaya, 2023; He et al., 2025).

*Corresponding author
ronnie.desouzasantos@ucalgary.ca (R.d.S. Santos); isantos3@hawaii.edu (I. Santos); matheus.demoraisleca@ucalgary.ca (M.d.M. Leça); cleyton.vanut@ufrpe.br (C. Magalhães); mairieli.wessel@ru.nl (M. Wessel)
ORCID(s): 0000-0003-3235-6530 (R.d.S. Santos)

Prior work indicates that LLMs contribute to multiple aspects of software development practice. Developers report improvements in productivity, reduced context switching, and faster onboarding, as these tools enable rapid access to examples, explanations, and partial solutions on demand (Fan et al., 2023; Coutinho et al., 2024; Rasnayaka et al., 2024). Time savings are particularly visible in documentation, bug fixing, and routine maintenance tasks (Hou et al., 2024; Gao et al., 2025; Haque, 2025). LLMs also support shorter feedback cycles between coding and testing activities, especially in early development stages, and assist developers in navigating unfamiliar codebases or implementation details (Sauvola et al., 2024; Chkirbene et al., 2024; Ozkaya, 2023; Haque, 2025; Santos et al., 2025). These capabilities suggest that LLMs are becoming an integral component of contemporary software engineering practice.

At the same time, a growing body of work has identified limitations and risks associated with LLM-assisted software engineering. Reported issues include biased or incorrect outputs, lack of transparency in how responses are generated, and inconsistencies across model versions (Sallou et al., 2024; Gao et al., 2025; He et al., 2025). As a result, LLM-generated artifacts often require careful human validation, particularly when dealing with complex logic, security-sensitive components, or domain-specific requirements (Ozkaya, 2023; Zheng et al., 2025; Haque, 2025). Additional concerns include hallucination, limited contextual awareness, variability in performance, and unresolved questions regarding code ownership and licensing (Sallou et al., 2024; Gao et al., 2025; Haque, 2025). These challenges indicate that, despite their usefulness, LLMs introduce new forms of uncertainty into software engineering practice.

Beyond these technical limitations, increasing attention has been directed toward how continuous interaction with LLMs may influence developer behavior and decision-making. In software engineering, early evidence suggests that extensive reliance on LLMs may reduce critical review, alter collaboration practices, and affect accountability during software development (Lo, 2023; Ozkaya, 2023; Marques et al., 2024; Chen and Alias, 2024; Haque, 2025). Parallel discussions in education and behavioral research provide additional context for these concerns. Studies suggest that frequent reliance on LLMs may reduce cognitive engagement, influence learning outcomes, and change how users process and retain information (Yankouskaya et al., 2024; Zhai et al., 2024; Kosmyna et al., 2025). In educational settings, extensive dependence has been associated with lower comprehension and reduced ownership of produced work (Zhai et al., 2024; Kosmyna et al., 2025). Behavioral research has also begun to examine functional dependence, emotional attachment, and difficulties moderating interactions with conversational AI systems (Yankouskaya et al., 2024; Ciudad-Fernández et al., 2025).

Although most software engineering studies have primarily focused on the effects of LLMs on productivity, code quality, correctness, and adoption, demonstrating that these tools can improve development efficiency and support a wide range of software engineering activities (Hou et al., 2024; Zheng et al., 2025; Haque, 2025; Ozkaya, 2023), comparatively little is known about how software engineers perceive their reliance on these systems during everyday development. As LLM-based tools are integrated into software engineering practice (Hou et al., 2024; Haque, 2025; Ozkaya, 2023), understanding productivity gains alone is insufficient, and understanding how software professionals perceive their reliance on these systems during everyday development has become an important topic. The broader literature on human-technology interaction distinguishes concepts such as dependence, overreliance, and addiction-related behaviors, each describing different forms of interaction with technology. Dependence may reflect the effective integration of LLMs into professional workflows (Fan et al., 2017; Zhang et al., 2024), whereas overreliance may reduce critical verification and appropriate calibration of trust in AI-generated outputs (Passi and Vorvoreanu, 2022; Klingbeil et al., 2024), and addiction-related behaviors may indicate difficulties moderating technology use (Marchica et al., 2022; Chamberlain et al., 2016). Understanding these distinctions is important for supporting responsible AI adoption, preserving developers' autonomy and critical judgment, and ensuring that increasing use of LLMs contributes to long-term professional development rather than replacing essential engineering expertise.

This study addresses this gap by exploring how software engineers experience the use of LLMs during professional software development, with particular attention to behavioral patterns associated with dependence, overreliance, and addiction-related behaviors. Rather than assessing clinical conditions, the study characterizes practitioners' experiences and perceptions of increasing reliance on these systems. To this end, this study addresses the following research question (RQ):

**RQ:** *How do software engineers experience the use of LLMs during professional software development, and what behavioral patterns associated with dependence, overreliance, and addiction-related behaviors emerge from their experiences?*

To answer this question, we conducted an exploratory qualitative survey with 119 software professionals working in industry across multiple countries. Our findings identified patterns consistent with functional dependence and overreliance in participants' reported use of LLMs. In contrast, behaviors associated with addiction-related issues are not frequent and primarily reflected perceived difficulty moderating LLM use or emotional attachment to the technology rather than compulsive behavior. In general, our research makes four contributions:

- Characterizes how software professionals experience and describe behavioral patterns associated with dependence, overreliance, and addiction-related behaviors during the daily use of LLMs.
- Provides empirical evidence from practicing software professionals, extending prior work that has primarily focused on students or general populations (Zhai et al., 2024; Zhou and Zhang, 2024).
- Provides a qualitative characterization of how LLMs are incorporated into professional software engineering workflows and how practitioners perceive the benefits and potential risks of their everyday use.
- Discusses implications for organizations, educators, and software practitioners seeking to promote responsible LLM adoption while preserving developers' autonomy, critical judgment, and well-being.

The remainder of this paper is organized as follows. Section 2 presents the conceptual background on dependence, overreliance, addiction-related behaviors, and LLM use in software engineering. Section 3 describes the research methodology. Section 4 presents the findings, which are discussed in Section 5. Finally, Section 6 concludes the paper.

## 2. Background

This section presents the conceptual foundations of this study. The first part defines dependence, overreliance, and addiction in the context of technology use. The second part characterizes the use of LLMs in software engineering and outlines their properties and limitations.

## 2.1. Dependence, Overreliance, and Addiction in Technology Use

Research on human interaction with technology has long addressed how repeated use of systems shapes user behavior, particularly in contexts such as decision support, automation, and interactive systems (Will, 1991; Fan et al., 2017; Baxter and Kabi, 2017). Within this body of work, dependence, overreliance, and addiction are treated as related but distinct constructs that capture different forms of engagement between individuals and technological artifacts (Passi and Vorvoreanu, 2022; Chamberlain et al., 2016; Alexander and Schweighofer, 1988). Table 1 summarizes these three concepts.

Dependence refers to the integration of a technology into routine practice due to its perceived usefulness, efficiency, or necessity (Fan et al., 2017). Empirical studies on technology use show that systems providing consistent task support tend to become embedded in everyday workflows, shaping how users approach and perform tasks (Fan et al., 2017; Will, 1991). In such cases, dependence reflects a functional adaptation, where users incorporate tools to reduce effort or improve performance. However, prior work also indicates that reliance on external systems may shift cognitive effort away from the user, potentially reducing direct engagement with underlying processes or diminishing opportunities for skill development (Zhang et al., 2024; Ali et al., 2024). This suggests that dependence is not inherently problematic, but its implications vary depending on context, task demands, and the degree of substitution between human and system capabilities.

Overreliance describes a different phenomenon, centered on how users evaluate and act upon system outputs rather than how frequently they use the system (Passi and Vorvoreanu, 2022). It is defined as a misalignment between user trust and system capability, where users place excessive confidence in the system and bypass independent judgment or verification (Klingbeil et al., 2024; Zhai et al., 2024). Research on automation and decision-making shows that users may follow system recommendations even when they are incorrect or when alternative information is available, a behavior associated with automation bias and overconfidence (Baxter and Kabi, 2017; Whyte and Sebenius, 1997). This pattern is often linked to cognitive processes such as heuristic reasoning and reduced analytical effort, where reliance on system outputs serves as a shortcut in complex decision contexts (Kahneman, 2011; Newell, 2013). As a result, overreliance is associated with decreased verification, reduced critical evaluation, and an increased likelihood of accepting incorrect outputs.

Addiction represents a more severe and qualitatively distinct form of engagement, characterized by compulsive use, diminished self-regulation, and continued interaction despite negative consequences (Chamberlain et al., 2016; Marchica et al., 2022). In the literature on behavioral addictions, technology use may become central in a user's life (salience), increase over time (tolerance), and produce negative psychological or functional effects when interrupted (withdrawal) (Marchica et al., 2022). Unlike dependence, which may remain functional, addiction implies a breakdown in control and the persistence of behavior even when it interferes with other responsibilities or well-being (Alexander and Schweighofer, 1988). Recent discussions have extended these frameworks to digital and AI systems, although the applicability and boundaries of addiction in these contexts remain debated (Yankouskaya et al., 2024; Ciudad-Fernández et al., 2025).

**Table 1**
Behavioral Characteristics of Dependence, Overreliance, and Addiction in Technology Use

| Characteristic | Dependence | Overreliance | Addiction |
|---|---|---|---|
| Definition | Functional, repeated use that may reduce independent thinking over time (Zhang et al., 2024; Fan et al., 2017). | Excessive trust beyond the system's capabilities (Will, 1991; Passi and Vorvoreanu, 2022). | Compulsive use despite negative consequences (Marchica et al., 2022; Chamberlain et al., 2016). |
| Voluntary Control | Mostly retained; may gradually decline with habitual use (Zhang et al., 2024). | Partially reduced due to misplaced confidence (Baxter and Kabi, 2017). | Impaired; behavior becomes difficult to regulate (Marchica et al., 2022). |
| User Awareness | Generally aware, although the extent of use may be underestimated (Ali et al., 2024). | Trust is perceived as justified despite system limitations (Passi and Vorvoreanu, 2022). | Often unaware of, or minimizes, negative consequences (Chamberlain et al., 2016). |
| Motivation | Utility, convenience, and routine (Fan et al., 2017). | Trust in system accuracy and reduced decision effort (Passi and Vorvoreanu, 2022). | Craving, emotional regulation, or compulsion (Marchica et al., 2022). |
| Impact on Behavior | May reduce critical thinking or autonomy (Zhang et al., 2024). | May lead to flawed decisions and overconfidence (Will, 1991). | May result in functional impairment, distress, or withdrawal symptoms (Marchica et al., 2022). |
| Reversibility | Usually reversible through intentional behavior change (Zhang et al., 2024). | Requires recalibration of trust and reassessment of system limitations (Baxter and Kabi, 2017). | May require behavioral or clinical support (Marchica et al., 2022). |
| Clinical Status | Not considered a clinical disorder (Ali et al., 2024). | Not clinically defined, although associated with practical and ethical concerns (Baxter and Kabi, 2017). | Recognized within behavioral addiction frameworks (e.g., DSM-5) (Marchica et al., 2022). |

These three constructs differ in their underlying mechanisms and implications. Dependence is primarily driven by perceived utility and task integration, overreliance by miscalibrated trust and insufficient verification, and addiction by compulsive engagement and reduced control (Passi and Vorvoreanu, 2022; Chamberlain et al., 2016). Although they may co-occur in practice, they capture distinct dimensions of human-technology interaction and therefore require separate consideration in empirical studies.

Recent work on AI systems suggests that these distinctions remain relevant as technologies become more capable, accessible, and embedded in everyday and professional activities (Zhou and Zhang, 2024; Zhai et al., 2024). Increasing system fluency, availability, and perceived competence introduce new conditions for reliance and trust, raising questions about how these forms of engagement manifest in interactions with contemporary AI tools.

### 2.2. Use of LLMs in Software Engineering

LLMs are deep learning systems trained on large textual corpora to generate natural language and code-related outputs (Hou et al., 2024; Zheng et al., 2025; Gao et al., 2025). Their ability to process and produce both natural language and source code has supported their adoption across a range of software engineering activities, where they act as interactive and generative tools in development workflows (Hou et al., 2024). Nowadays, these tools are used across multiple stages of the software development lifecycle, including requirements analysis, code generation, program repair, testing, and maintenance (Hou et al., 2024; Zheng et al., 2025). Their use is particularly visible in implementation-related activities, where they assist with coding, debugging, and documentation (Haque, 2025; Ozkaya, 2023). In practice, developers often use LLMs to scaffold initial solutions, resolve syntax-related issues, and support debugging, especially when working with unfamiliar technologies or in early development phases (Rasnayaka et al., 2024). These forms of use suggest that LLMs are incorporated into everyday development practices as tools that complement existing skills and workflows.

At the same time, the literature describes several limitations associated with LLM-based support. Generated outputs may be syntactically correct while containing semantic errors or inconsistencies, reflecting the probabilistic nature of these models (Ozkaya, 2023; Haque, 2025). LLMs may also produce plausible but incorrect solutions, which can introduce faults into software artifacts when not carefully validated (Fan et al., 2023). In addition, outputs may vary across executions, as identical prompts can produce different results, which introduces challenges for reproducibility and systematic evaluation (Fan et al., 2023; Zheng et al., 2025). These characteristics demonstrate that, while LLMs can support development tasks, their outputs require interpretation and verification by developers.

The combination of broad applicability, ease of interaction, and imperfect reliability influences how LLMs are used in practice. Their capacity to reduce effort and provide rapid feedback supports their integration into routine workflows, which aligns with patterns of dependence observed in other forms of technology use (Fan et al., 2017; Rasnayaka et al., 2024). At the same time, the fluency and apparent coherence of generated outputs may affect how users assess their correctness, creating conditions in which outputs are accepted without sufficient verification, a pattern consistent with overreliance (Passi and Vorvoreanu, 2022; Klingbeil et al., 2024; Zhai et al., 2024). Discussions in recent work also raise questions about sustained and intensive engagement with AI systems, including whether repeated and reinforcing interactions may extend toward patterns associated with addiction, although such interpretations remain debated and require further empirical clarification (Yankouskaya et al., 2024; Ciudad-Fernández et al., 2025; Haman and Školník, 2023). These observations suggest that the use of LLMs in software engineering can be understood not only in terms of technical capabilities, but also in relation to broader patterns of human-technology interaction.

## 3. Method

This study employed a qualitative survey to explore how software engineers experience the use of LLMs during professional software development and the behavioral patterns associated with dependence, overreliance, and addiction-related behaviors. Qualitative surveys are an established method in software engineering for investigating practitioners' experiences, particularly when the objective is to understand an emerging phenomenon rather than estimate its prevalence Melegati et al. (2024). Unlike quantitative surveys, which seek statistical descriptions of a population by measuring the distribution of predefined variables or testing relationships among them, qualitative surveys examine the diversity of experiences, perspectives, and meanings within a target population Braun et al. (2021). They may include structured questions, such as Likert scales, multiple choice items, and demographic questions, but these responses are interpreted conceptually rather than analyzed as variables for statistical inference. Consequently, the unit of analysis is the meaning conveyed by participants' responses and the patterns that emerge across them, making qualitative surveys particularly appropriate for exploratory research on socio-technical phenomena in software engineering.

This method is appropriate because the behavioral implications of LLM use among software engineers remain insufficiently understood. Rather than estimating the prevalence of dependence, overreliance, or addiction-related behaviors, our objective was to explore how software engineers experience LLM use in their professional activities and identify the behavioral patterns associated with these experiences. To achieve this, we designed an anonymous online qualitative survey targeting software engineers working in professional software development. The survey combined demographic questions, structured questions using Likert scales and categorical response options, and open-ended questions addressing perceived reliance, changes in work practices, cognitive engagement, verification behaviors, emotional attachment, and self-regulation. The questionnaire was informed by prior research on technology dependence, automation overreliance, and behavioral addiction (Zhang et al., 2024; Fan et al., 2017; Ali et al., 2024; Will, 1991; Baxter and Kabi, 2017; Passi and Vorvoreanu, 2022; Marchica et al., 2022; Chamberlain et al., 2016), which guided the development of questions related to perceived control, trust, cognitive engagement, emotional regulation, and substitution of human judgment.

### 3.1. Questionnaire Design

The survey questionnaire was designed following established guidelines for survey research in software engineering, including iterative item development, alignment between research objectives and survey questions, and refinement through discussions among the research team (Linaker et al., 2015; Ralph et al., 2020). The overall design also followed the methodological guidance on qualitative surveys, where structured and open-ended questions are combined to capture the diversity of practitioners' experiences rather than produce statistical estimates (Melegati et al., 2024; Braun et al., 2021).

The questionnaire was developed from the conceptual framework presented in Section 2, where dependence, overreliance, and addiction are treated as related but distinct constructs. Rather than adapting an existing validated instrument, we operationalized these constructs by deriving questions directly from the behavioral characteristics identified in the literature:

- **Dependence (Q1–Q5):** These items were informed by research describing dependence as the functional integration of technology into everyday work through perceived usefulness, habitual use, and cognitive engagement (Fan et al., 2017; Zhang et al., 2024; Ali et al., 2024). They investigated cognitive preoccupation, increasing integration of LLMs into work routines, reflection on previous interactions, and the use of LLMs to manage uncertainty or maintain focus during software development.

- **Overreliance (Q6–Q8):** These items were based on studies defining overreliance as excessive trust in automated systems and insufficient verification of their outputs (Will, 1991; Baxter and Kabi, 2017; Passi and Vorvoreanu, 2022; Klingbeil et al., 2024). They explored participants' willingness to rely on LLM-generated code without human review, preference for LLMs over alternative information sources, and the substitution of collaboration or peer consultation with AI-generated guidance.

- **Addiction Related Behaviors (Q9–Q18):** These items were informed by the behavioral addiction literature, particularly work describing salience, tolerance, withdrawal, impaired self-regulation, emotional regulation, and functional impairment associated with excessive technology use (Marchica et al., 2022; Chamberlain et al., 2016; Alexander and Schweighofer, 1988). They investigated behaviors such as spending more time than intended using LLMs, increasing urges to use them, unsuccessful attempts to reduce usage, emotional discomfort when access was unavailable, and perceived negative consequences of excessive use.

In addition to these construct-specific questions, the questionnaire included two contextual open-ended questions asking participants to describe situations in which they used LLMs during software development and how their work would be affected if such tools were unavailable. These questions provided contextual information to support the interpretation of the structured responses and allowed participants to describe experiences not captured by predefined response options. The final questionnaire consisted of demographic questions, contextual questions, structured questions using five-point Likert scales, categorical questions, and open-ended questions (Table 2). For the behavioral questions, response options ranged from 1 ("Never") to 5 ("Very Often"), while the overreliance question regarding production code adoption used categorical response options. Demographic information, including participants' professional role, years of experience, country of residence, prior AI training, and the types of software systems they developed, was collected to characterize the participant pool and contextualize the findings. As the purpose of this qualitative survey was to explore the diversity of practitioners' experiences rather than estimate population characteristics or test statistical relationships, these variables were used descriptively and not for inferential analysis. To reduce potential ordering effects, the presentation of the behavioral questions was randomized (Linaker et al., 2015). The survey was implemented using Qualtrics.[1]

### 3.2. Pilot and Questionnaire Refinement

The questionnaire underwent an iterative refinement process before data collection to improve clarity, content coverage, and appropriateness for the study objectives. Rather than relying on a single pilot, the instrument was progressively refined each time feedback was received from individuals with complementary expertise. In this process, the first iteration involved two practicing software engineers, who completed the questionnaire and provided feedback on question wording, clarity, completion time, and the overall flow of the survey. Their comments helped identify ambiguous wording, improve readability, and ensure that the instrument was appropriate for practitioners working in professional software development.

Following these revisions, two researchers with expertise in artificial intelligence and software engineering reviewed the questionnaire. This stage focused on assessing

[1] https://www.qualtrics.com

**Table 2**
Survey Questions

| Question |
|---|
| **Context** |
| **QC1.** Describe a specific situation when you used an LLM (AI tool) to assist with a coding or programming task at work. |
| **QC2.** Imagine that you arrive at work one day and do not have access to the current LLM tools (e.g., ChatGPT, Copilot). How would that affect your daily tasks? |
| **Dependence** |
| **Q1.** How often have you spent time thinking about using LLM tools (e.g., ChatGPT, Copilot) in your work tasks? |
| **Q2.** How often have you planned how to increase the use of LLMs (e.g., ChatGPT, Copilot) in your workday tasks? |
| **Q3.** How often have you reflected on previous interactions you had with LLMs in your work? |
| **Q4.** How often have you used LLMs to reduce feelings of uncertainty or helplessness when stuck at work? |
| **Q5.** How often have you used LLMs in order to reduce restlessness? |
| **Overreliance** |
| **Q6.** Would you be comfortable using LLM-generated code in production systems without human review? |
| **Q7.** How often have you prioritized LLM use over other resources (e.g., documentation, peer discussions)? |
| **Q8.** How often have you ignored coworkers or meetings and trusted LLMs to complete your work tasks? |
| **Addiction Related Behaviors** |
| **Q9.** How often have you spent more time using LLMs in your tasks than you initially intended? |
| **Q10.** How often have you felt an increasing urge to use LLMs more in your tasks? |
| **Q11.** How often have you needed to use LLMs for longer periods to get the same sense of help or satisfaction? |
| **Q12.** How often have you used LLMs to cope with feelings of stress or anxiety about your work? |
| **Q13.** How often have you been told by others (colleagues, mentors) to reduce your LLM use but not followed that advice? |
| **Q14.** How often have you tried to reduce your use of LLMs in your work without success? |
| **Q15.** How often have you planned to rely less on LLMs in your work but were unable to follow through? |
| **Q16.** How often have you felt restless or uneasy when you could not use an LLM while doing your work? |
| **Q17.** How often have you become irritable when LLMs were unavailable (e.g., internet down, system blocked) while doing your work? |
| **Q18.** How often have you used LLMs so much that it negatively affected your job performance or work tasks? |

*Note:* Q1–Q5 and Q7–Q18 used a five point frequency scale ranging from 1 ("Never") to 5 ("Very Often"). Q6 used categorical response options: Yes, No, and Only for non critical tasks.

whether the questions were appropriate, understandable, and relevant to professional software engineering practice, and whether the scenarios and terminology accurately reflected the use of LLMs in industrial development. Based on their feedback, we improved clarity and reduced overlap between items. Given the behavioral nature of the study, the questionnaire was subsequently reviewed by a licensed psychologist, who is also an academic professor, with expertise in Cognitive Behavioral Therapy, hospital psychology, and psychology education. This review focused on ensuring that the instrument explored participants' experiences and perceptions without suggesting clinical assessment or diagnosis. Based on this feedback, additional refinements were made to avoid clinical language and ensure that the instrument remained appropriate for investigating technology-related behaviors rather than mental health conditions.

After each review stage, the research team discussed the feedback and revised the questionnaire accordingly. The iterative refinement process improved the clarity, relevance, and appropriateness of the questions, while ensuring that they adequately represented the conceptual dimensions derived from the literature and reflected the context of professional software engineering. Because this study employed an exploratory qualitative survey rather than a psychometric instrument, it did not seek to establish construct validity through statistical validation procedures. Instead, the questionnaire was developed by operationalizing the theoretical constructs of dependence, overreliance, and addiction-related behaviors identified in prior work. The iterative review process provided indirect support for this operationalization by evaluating whether the questions were understandable, appropriate for software engineering practice, and consistent with the intended non-clinical interpretation of the behavioral constructs. Consequently, these activities should not be interpreted as formal psychometric validation of the instrument. In parallel, technical testing was conducted across multiple browsers and devices to verify the usability and stability of the Qualtrics implementation. Responses collected during these refinement activities were excluded from the final dataset.

## 3.3. Sampling and Recruitment

Participants were mainly recruited using purposive and convenience sampling (Baltes and Ralph, 2022). The target population consisted of software practitioners who self-reported current or previous experience working in software engineering and using LLMs during professional software development. No restrictions were imposed regarding age, employment status, or geographic location, provided participants satisfied the study eligibility criteria. The primary recruitment channel was Prolific, an online participant recruitment platform widely used in empirical software engineering research because it provides access to diverse participant pools and built-in mechanisms for participant screening and quality control (Reid et al., 2022; Russo, 2022). To maximize diversity of professional experiences, participant recruitment was stratified across continents. Eligibility criteria available through Prolific were configured to recruit participants working in software engineering or related computing activities. The survey link was also disseminated through professional mailing lists to broaden participation among software practitioners and enable snowball sampling (Baltes and Ralph, 2022).

Although Prolific provides access to a large and geographically diverse pool of participants, previous research has shown that its prescreening criteria rely primarily on self-reported information, which may not accurately reflect participants' technical skills or professional experience (Reid et al., 2022; Russo, 2022; Alami et al., 2024).

Consequently, relying exclusively on the platform's recruitment mechanisms may result in the inclusion of participants who do not satisfy the intended target population. To improve the validity of the sampling process, respondents were required to satisfy predefined eligibility criteria, and the survey incorporated additional screening and attention check questions to verify participants' suitability for the study. The application of these eligibility and quality criteria during data cleaning is described in Section 3.4.

Because the survey was disseminated through both Prolific and professional mailing lists using the same survey link, it was not possible to distinguish participants by recruitment source or calculate the response rate. The final dataset consisted of 119 valid responses. However, consistent with the objectives of a qualitative survey (Braun et al., 2021; Melegati et al., 2024), our sampling strategy was intended to recruit software practitioners with relevant experience rather than to obtain a statistically representative sample of the software engineering population.

### 3.4. Filtering and Data Quality Control

Following data collection, responses were filtered according to the predefined eligibility and quality criteria established during survey design. First, the eligibility criteria were verified through embedded screening questions that assessed participants' current role, involvement in software engineering tasks, and familiarity with LLM tools. Responses from participants who did not satisfy the study criteria were excluded. Second, consistency checks were applied to identify potentially low-quality responses. These included detecting contradictions between related answers, unusually short completion times, and patterned responses indicative of inattentive answering. Responses failing these consistency checks were excluded from the dataset. Third, incomplete submissions were removed to ensure that all analyzed responses contained sufficient information for interpretation. These filtering procedures follow established practices for improving data quality in survey-based software engineering research (Danilova et al., 2021; Alami et al., 2024) and increase confidence that the analyzed dataset reflects responses from participants who met the study criteria and provided sufficiently complete and consistent answers.

### 3.5. Data Analysis

Consistent with the qualitative survey methodology adopted in this study, the objective of the analysis was to characterize how software engineers experience the use of LLMs during professional software development rather than to estimate the prevalence of specific behaviors or test statistical relationships among variables. Therefore, we analyzed the questionnaire as a whole, with both structured and open-ended responses contributing complementary evidence toward understanding the phenomenon.

The structured questions, including Likert scale and categorical items, were analyzed using descriptive statistics (George and Mallery, 2018). Frequencies, percentages, medians, means, and standard deviations were calculated to summarize participants' responses across the three conceptual dimensions investigated: dependence, overreliance, and addiction-related behaviors. In keeping with the objectives of qualitative surveys (Braun et al., 2021; Melegati et al., 2024), these descriptive results were not interpreted as estimates of population prevalence or used to test statistical hypotheses. Instead, they were used to support the characterization of behavioral patterns emerging from participants' reported experiences and to complement the qualitative interpretation of the data.

We analyzed the open-ended questions using reflexive thematic analysis (Cruzes and Dyba, 2011; Terry et al., 2017). This approach was selected because it supports the identification and interpretation of recurring patterns across participants' experiences while preserving the contextual meaning of their narratives. The analysis followed four stages, illustrated in Figure 1:

- **Familiarization**: Two researchers independently read all responses several times to become familiar with the data and record initial observations.
- **Open coding**: The responses were manually coded using an inductive approach. Codes were developed directly from participants' accounts, emphasizing their own descriptions and interpretations rather than predefined categories.
- **Code reconciliation**: The two researchers compared and discussed their coding decisions, resolving disagreements through discussion and refining the coding scheme until consensus was reached. A third researcher was available to mediate disagreements when necessary.
- **Theme development**: Related codes were iteratively grouped into higher-level themes representing recurring experiences associated with LLM use during professional software development.

Shared coding spreadsheets were used to document coding decisions and support transparency throughout the analysis. Automated coding tools were intentionally not employed because preserving the context and nuance of participants' responses was considered essential. It is important to note that very brief responses that did not contain sufficient interpretive content (e.g., “Quite a lot” or “Greatly”) were coded as *N/A* and excluded from thematic development. We concluded the analysis when additional iterations no longer resulted in new codes or substantive refinements to the thematic structure, indicating adequate thematic coverage for the purposes of this exploratory study (Ralph et al., 2020). At the end, the findings from the structured and open-ended questions were interpreted together to develop a richer characterization of software engineers' experiences with LLM use and the behavioral patterns associated with dependence, overreliance, and addiction-related behaviors.

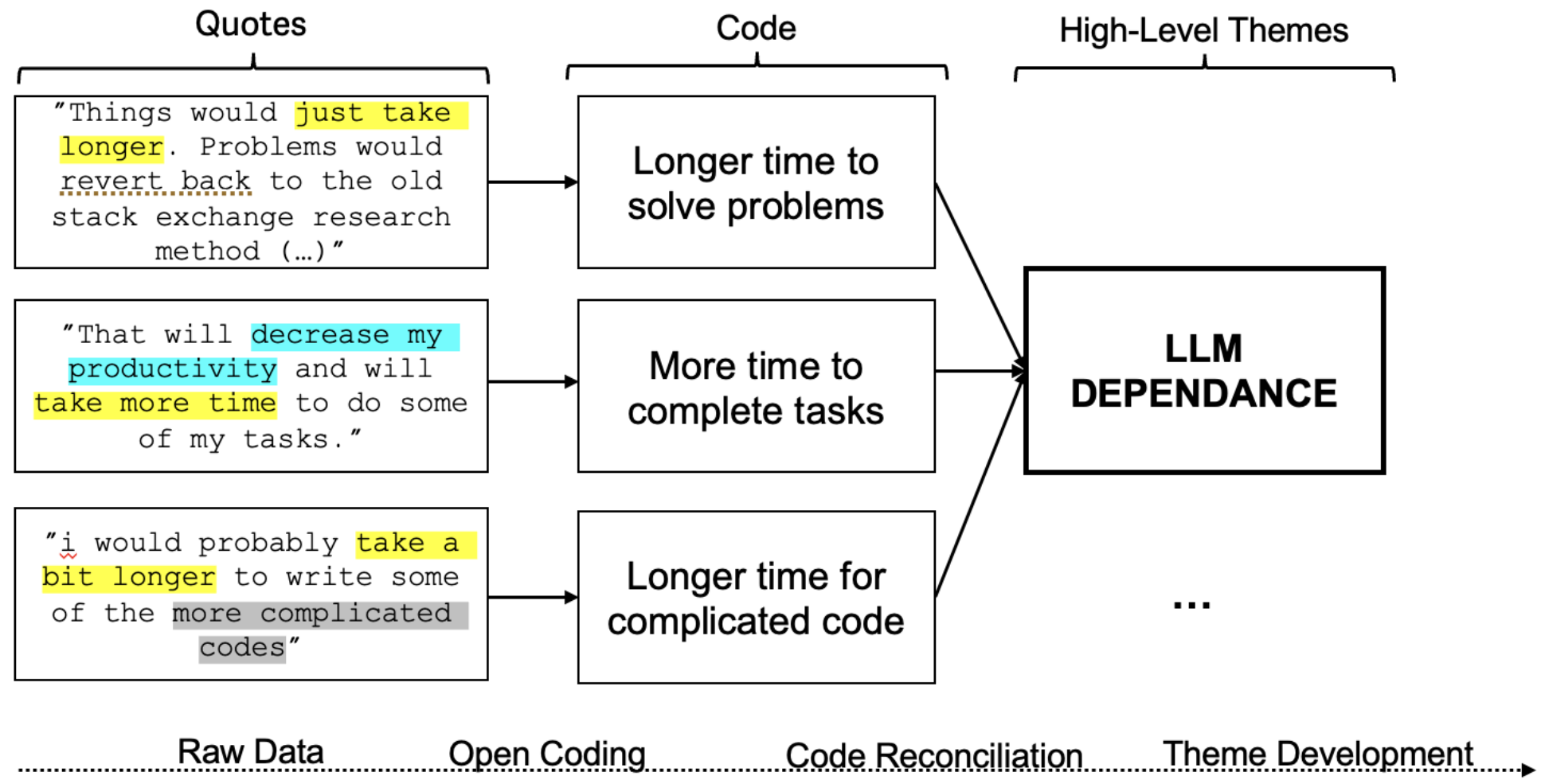

**Figure 1:** Thematic Analysis

### 3.6. Ethics

This study was conducted in accordance with the institutional ethics guidelines of the first author's university and received approval from its Research Ethics Board. Participation was voluntary, and informed consent was obtained from all participants before they completed the survey. The questionnaire was anonymous, and no personally identifiable information was collected or stored. Participants were informed that they could withdraw from the study at any time before submitting their responses.

### 3.7. Positionality Statement

All authors of this study conduct research on human aspects of software engineering, with a particular focus on behavior, collaboration, and interaction with software technologies. Their research background informed the design of this study, including the selection of constructs related to dependence, overreliance, and behavioral patterns associated with LLM use. The authors also have prior experience working with software development teams in both academic and industry contexts. This experience contributes to their understanding of real-world development practices, including how tools are integrated into workflows and how practitioners make decisions under time and resource constraints. From a methodological perspective, the authors are trained in empirical software engineering and have experience with both quantitative and qualitative methods. This background influenced choices related to survey design, data interpretation, and the integration of multiple data sources. In particular, familiarity with thematic analysis and behavioral constructs may have guided the identification and interpretation of patterns in the qualitative data. To mitigate potential bias, the study was conducted collaboratively, with multiple authors involved in instrument design, data analysis, and interpretation. This collaborative process enabled cross-checking of interpretations and discussion of alternative explanations. Consistent with prior discussions on positionality in software engineering research (De Sousa et al., 2025), the authors acknowledge that their professional and research backgrounds form part of the analytical lens through which the data are interpreted and that complete neutrality is neither expected nor assumed.

### 3.8. Threats to Validity

Consistent with the established guidelines for reporting survey-based research Kitchenham and Pfleeger (2008); Ralph et al. (2020); Braun et al. (2021); Melegati et al. (2024), we discuss the principal threats to validity that arise from the adopted methodology and should be considered when interpreting the findings.

**Construct validity**. The operationalization of dependence, overreliance, and addiction-related behaviors was informed by prior literature. Nevertheless, these constructs remain inherently subjective and context-dependent. To reduce this threat, survey items were derived from existing conceptualizations, refined through pilot testing, and reviewed by the research team. The qualitative analysis prioritized participants' original wording and interpreted responses within their reported context rather than imposing predefined classifications.

**Internal validity**. The study relies on self-reported perceptions of LLM use, which may be affected by recall bias, social desirability bias, or differences in participants' interpretation of survey questions. To reduce these threats, the questionnaire combined closed and open-ended questions, allowing qualitative responses to contextualize and clarify participants' quantitative answers. The analysis emphasized recurring patterns across participants rather than isolated responses.

**Reliability**. The qualitative analysis involves interpretive judgment and is therefore susceptible to researcher bias. To improve consistency, two researchers independently coded the qualitative responses, followed by discussion and reconciliation of disagreements until consensus was reached.

Coding decisions and theme definitions were iteratively refined throughout the analysis.

**External validity**. Participants were recruited primarily through Prolific and complemented by professional mailing lists. Although this strategy enabled participation from software practitioners with diverse geographic locations, roles, and experience levels, the resulting sample is not statistically representative of the broader software engineering population. Consequently, the findings should not be interpreted as statistically generalizable. Instead, they provide evidence of patterns that may transfer to similar software engineering contexts.

**Sampling and data quality**. Previous research has shown that online recruitment platforms relying on self-reported participant characteristics may include individuals who do not fully satisfy the intended target population. To mitigate this threat, we combined platform-level prescreening with study-specific eligibility verification, attention checks, and post-collection filtering procedures, as described in Sections 3.3 and 3.4. These procedures increase confidence that the analyzed dataset reflects responses from participants who satisfied the study criteria.

**Generalizability**. The objective of this study was to understand practitioners' perceptions and experiences rather than to estimate population parameters. Therefore, our findings should be interpreted as exploratory. Future studies employing alternative sampling strategies, larger samples, longitudinal designs, or complementary empirical methods may further investigate the prevalence and evolution of the identified behaviors.

## 4. Results

This section presents the findings from the qualitative survey. We begin by describing the characteristics of the participants. We then present the findings organized around the three behavioral factors investigated in this study: dependence, overreliance, and addiction. For each factor, descriptive statistics summarize participants' responses to the closed-ended questions, followed by the resulting analysis of the open-ended responses. Finally, we integrate these findings to answer the research question.

### 4.1. Participant Demographics

Our final dataset consisted of 119 software professionals with diverse backgrounds, representing different levels of professional experience, software engineering roles, geographic regions, and formal AI training. Consistent with our purposive sampling strategy, our objective was not to obtain a statistically representative sample but to capture a broad range of professional experiences and perspectives regarding the use of LLMs in software engineering.

We recruited participants from 17 countries. The largest groups were from the United States (19.3%, $n = 23$), the United Kingdom (15.1%, $n = 18$), South Africa (13.4%, $n = 16$), and Canada (11.8%, $n = 14$). Additional participants were from Australia (6.7%, $n = 8$), Spain (5.9%, $n = 7$), India (4.2%, $n = 5$), and Germany, Chile, Brazil, and New Zealand (3.4%, $n = 4$ each). Mexico and Japan each contributed three participants (2.5%), while Italy, Portugal, Ireland, and France were represented by one or two participants. This geographic diversity allowed us to capture experiences from practitioners working in different industrial, organizational, and cultural contexts, reducing the likelihood that the findings reflect practices from a single software ecosystem.

Our sample also included professionals with a wide range of experience, from less than one year to more than ten years. The largest group reported 1–3 years of experience (29.4%, $n = 35$), followed by more than 10 years (23.5%, $n = 28$), 3–5 years (21.0%, $n = 25$), 5–10 years (20.2%, $n = 24$), and less than one year (5.9%, $n = 7$). This distribution enabled us to capture perspectives from both relatively early career and experienced practitioners, allowing the analysis to reflect a range of levels of familiarity with software engineering practices.

Participants also represented a variety of software engineering roles, including Data Scientists (28.6%, $n = 34$), Full Stack Developers (21.8%, $n = 26$), Frontend Developers (12.6%, $n = 15$), Backend Developers (10.9%, $n = 13$), DevOps Engineers (8.4%, $n = 10$), and Quality Assurance professionals (2.5%, $n = 3$). The remaining participants (15.1%, $n = 18$) reported roles such as software architects, requirements engineers, engineering managers, and technical leads. Including professionals performing different software engineering activities supported the exploration of LLM use across a range of technical responsibilities rather than within a single specialization.

Finally, 37.8% ($n = 45$) reported formal training in AI or LLMs, whereas 62.2% ($n = 74$) had not received formal training. This variation allowed us to explore experiences from practitioners with different levels of formal preparation in AI, rather than focusing exclusively on AI specialists. Regarding gender, 65.5% ($n = 78$) identified as men, 33.6% ($n = 40$) as women, and one participant (0.8%) preferred not to disclose their gender. Although gender was not used as an analytical variable, reporting this information provides additional context for interpreting the diversity of the participant pool.

### 4.2. Software Engineers and LLMs: Dependence, Overreliance, and Addiction

Participants reported behaviors associated with dependence, overreliance, and addiction-related behaviors, although the frequency and nature of these behaviors differed across the three categories. In general, our findings suggest that while LLM use is commonly integrated into participants' daily activities, the reported behaviors reflect different ways of engaging with these tools. The following sections present the findings for each behavior separately, combining the quantitative results with participants' qualitative accounts to provide a more comprehensive characterization of their experiences. The figures report the distribution of responses to the Likert scale questions and compare these patterns across different levels of professional experience.

Lower-frequency responses are shown on the left (yellow), neutral responses in the center (gray), and higher-frequency responses on the right (green).

#### *4.2.1. Dependence*

Across the five questions (Q1–Q5) shown in Figure 2, a clear pattern emerged: the majority of participants reported engaging often or very often in thoughts and behaviors indicative of habitual use. Most respondents frequently thought about using LLMs during their work, reported planning to increase LLM use in their tasks, and reflected on prior interactions with these tools. These results show that LLM use is frequently reported across planning, reflection, and task-related activities. Emotional aspects of use were also observed: a majority of participants reported using LLMs when facing uncertainty, and a similar majority reported using them in situations associated with restlessness, a pattern that held consistently across all experience levels. The similarity in response distributions across Q1–Q5 indicates that frequent use is reported across multiple situations rather than being concentrated in a single type of activity.

When separating responses by experience level (Figure 3), mid-level professionals consistently reported the highest frequency of “Often” and “Very Often” responses across the first three questions, ahead of both entry-level and senior cohorts. This pattern held for thinking about using LLMs (Q1), planning to increase LLM use (Q2), and reflecting on past interactions (Q3), where the mid-level subset reported markedly more frequent engagement than either the entry-level or senior subgroups. The emotional items showed a somewhat different distribution across cohorts. For using LLMs when facing uncertainty (Q4), the mid-level group again reported the highest frequency, while entry-level professionals reported the lowest. For restlessness-related use (Q5), however, the pattern reversed: senior professionals reported the highest frequency of all three cohorts, followed by mid-level and then entry-level respondents, making this the only item in the set where the senior subgroup led rather than the mid-level subgroup. Overall, frequent use was reported across all experience levels, with mid-level professionals showing the strongest concentration for most items and senior professionals standing out specifically for restlessness-related use. This pattern indicates that reported use spans both task-related and situational contexts rather than being limited to a specific type of activity or experience group.

These patterns are reflected in how participants described their work when LLMs were unavailable. Forty five participants reported slower task completion, increased effort, or reduced efficiency in the absence of these tools. For example, one participant explained, *“That will decrease my productivity and will take more time to do some of my tasks” (P004)*, while another stated, *“I’ll have a less efficient time at work, having to find everything by myself taking much more time” (P009)*. Similar responses referred to longer work duration, additional steps in problem solving, and increased reliance on manual search. These accounts provide context for the frequent dependence-related behaviors reported in the survey by showing that participants primarily associated LLM dependence with improving the efficiency of routine software engineering activities. Rather than preventing task completion, the absence of LLMs was generally described as making activities slower and more effortful.

#### *4.2.2. Overreliance*

One key indicator of overreliance is trust, specifically whether software engineers report willingness to adopt LLM outputs without additional review (Q6). In this question, a minority of participants stated they would be comfortable using LLM-generated code in production systems without human review, while the majority said they would not, and a smaller group indicated they would do so only for non-critical tasks. These responses indicate that most participants report some level of caution when directly using generated outputs. In contrast, responses to Q7 and Q8 show higher frequencies of use. In Q7, roughly half of participants reported “Often” or “Very Often” prioritizing LLMs over other resources such as documentation or peer discussions. Similarly, in Q8, a clear majority reported frequently relying on LLMs instead of engaging with coworkers or meetings. These results suggest that although participants generally report caution when directly accepting LLM outputs, many also describe routinely turning to these tools as their primary source of information and support.

When analyzing responses by experience level (Figure 5), the two behaviors showed different patterns across cohorts. In Q7, entry-level and mid-level professionals reported comparable, elevated frequencies of prioritizing LLMs over documentation or peer discussion, while senior professionals reported this behavior noticeably less often than either of the other two cohorts. In Q8, by contrast, mid-level professionals reported the highest frequency of relying on LLMs instead of engaging with coworkers or meetings, followed by entry-level and then senior professionals, who reported comparatively similar and lower frequencies to one another. These results indicate that while caution in directly adopting LLM-generated outputs is reported across experience levels, frequent reliance on LLMs as a source of support is also common, particularly among early and mid-career professionals. This contrast suggests that reported trust in LLM outputs and reliance on LLMs during everyday work do not necessarily occur together, and that the specific way this reliance is expressed differs by career stage.

Participants’ reports of how they would respond to the absence of LLMs provide additional context for these findings. Nineteen participants described behaviors associated with overreliance, particularly the tendency to treat LLMs as their preferred source of support. Some indicated that they would delay work until access to an LLM was restored or described these tools as their default approach to solving development problems. For example, P115 stated, *“It would not affect me that much. I would wait until the LLM was available again,”* while P050 commented, *“It helps me to*

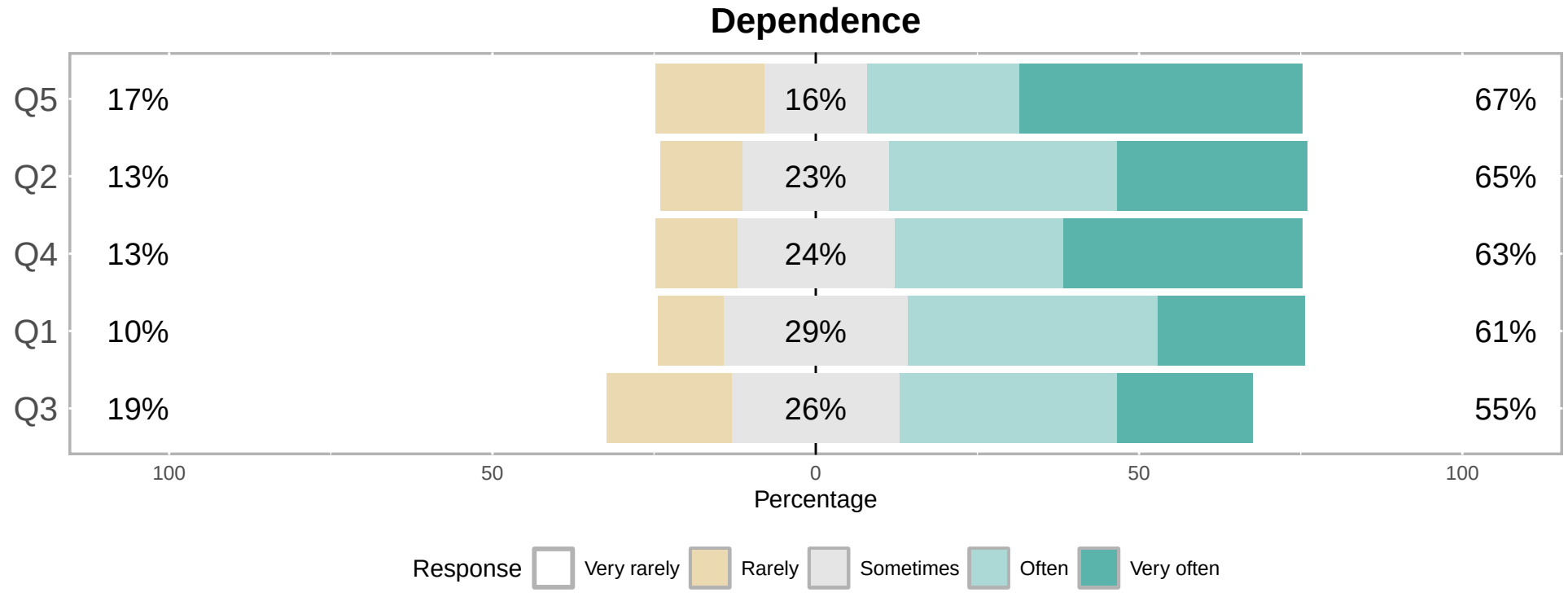

**Figure 2:** Responses to Likert-scale questions – dependence.

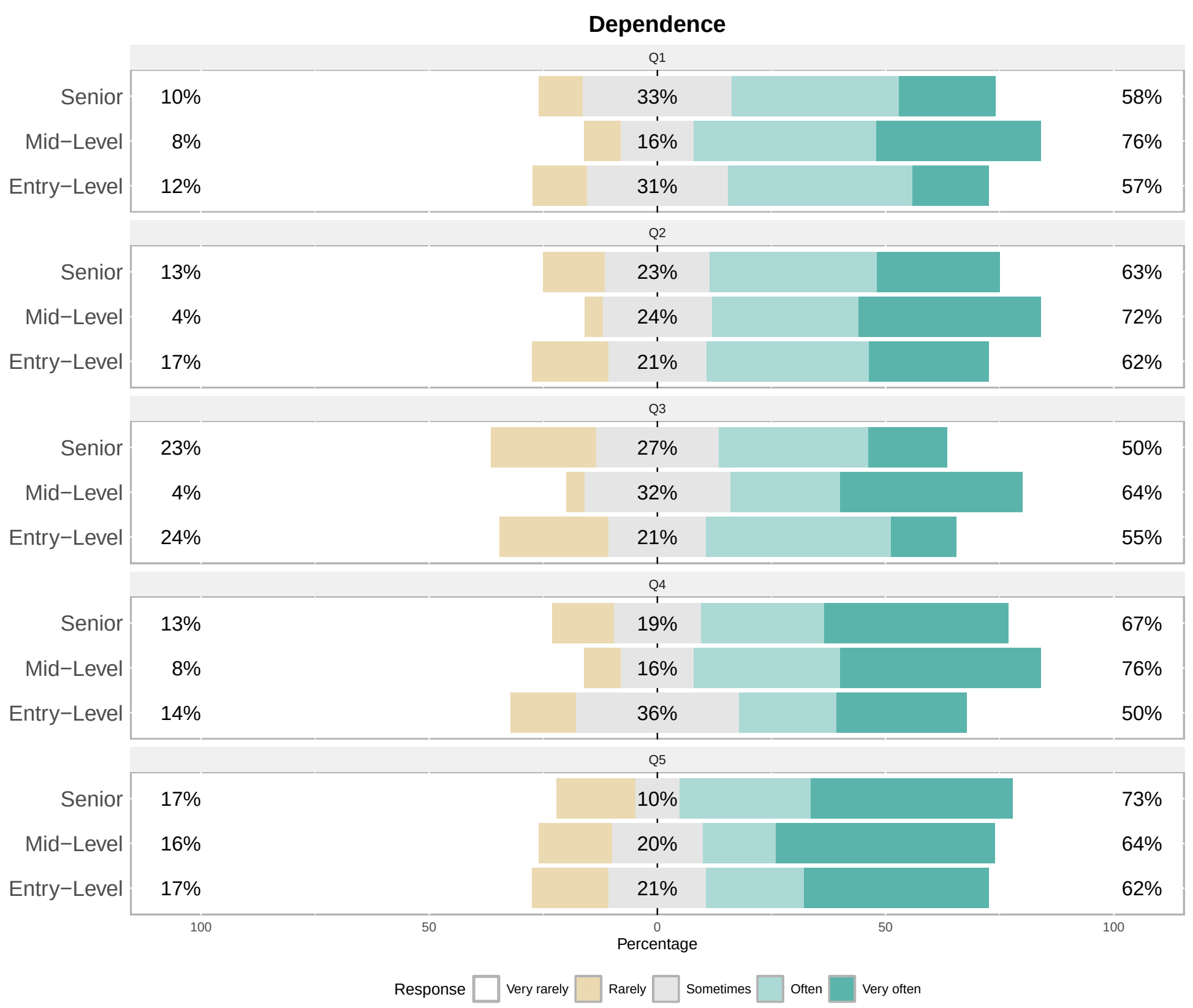

**Figure 3:** Responses to Likert-scale questions by experience level – dependence.

*do so much at work, I can't possibly imagine a day without it."* Other participants reported that they would only consult colleagues or documentation when LLMs were unavailable, suggesting that alternative sources of information had become secondary rather than complementary resources. These experiences indicate that overreliance is reflected primarily in participants' tendency to position LLMs as their preferred source of support during software development, with alternative resources becoming secondary rather than complementary.

### *4.2.3. Addiction*

The main characteristic associated with addiction-related behaviors is difficulty regulating the use of a technology, particularly when this use extends beyond its intended purpose or becomes associated with emotional responses. Across the ten questions (Q9–Q18) shown in Figure 6, participants frequently reported behaviors related to time use, difficulty reducing use, and reactions when access to LLMs was unavailable. For most items, more than half of participants selected "Often" or "Very Often." High frequencies were observed across multiple items, including spending more time than intended using LLMs (Q9), planning to rely less on LLMs but being unable to do so (Q15), and feeling restless when unable to use these tools (Q16). Similarly frequent responses were observed for using LLMs in situations associated with stress (Q12) and becoming irritable when access was unavailable (Q17). These distributions were relatively consistent across the questions, suggesting that these behaviors were reported across multiple aspects of LLM use rather than being concentrated in a single behavior.

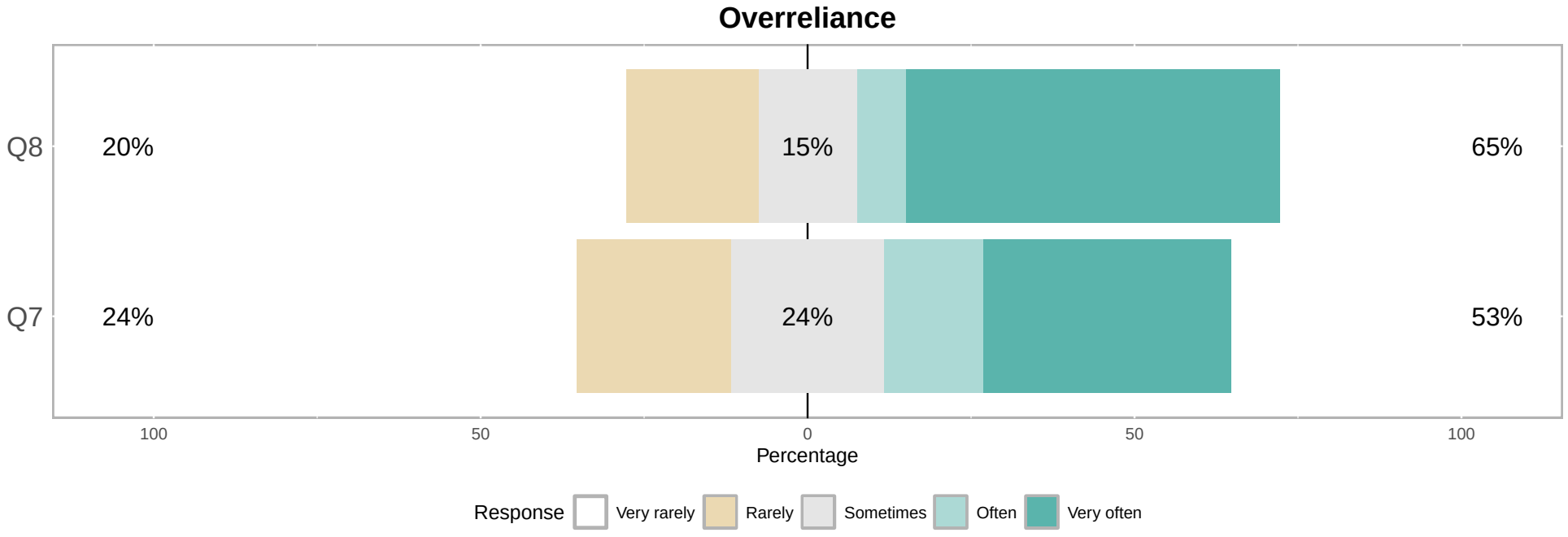

**Figure 4:** Responses to Likert-scale questions – overreliance.

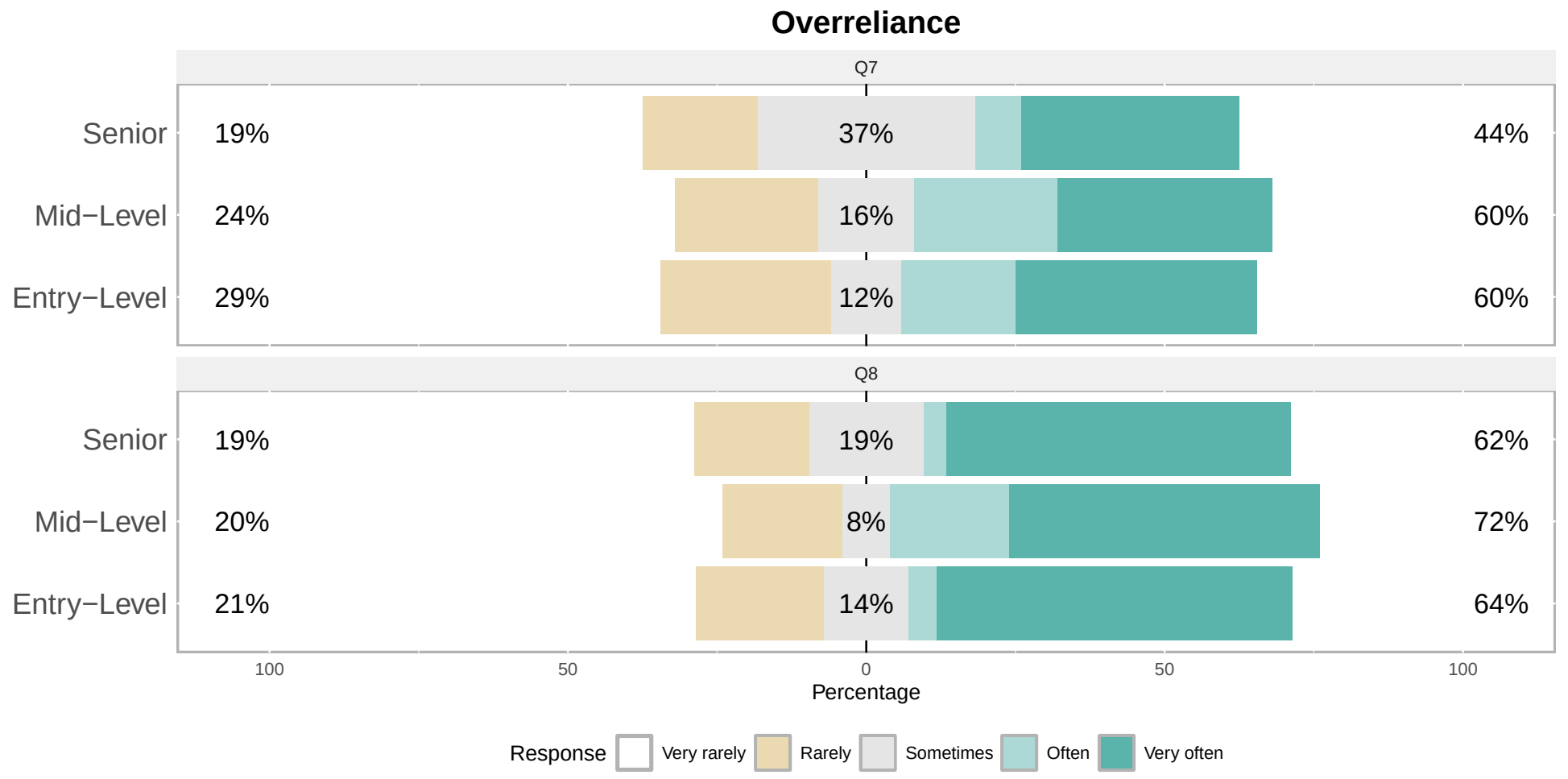

**Figure 5:** Responses to Likert-scale questions by experience level – overreliance.

When responses were analyzed by experience level (Figure 7), the pattern varied by behavior type rather than following a single trend across cohorts. Mid-level professionals reported the highest frequencies on items related to time spent using LLMs and difficulty reducing their use, ahead of both entry-level and senior cohorts. Entry-level professionals, in turn, reported the highest frequencies on items related to emotional reactions, such as restlessness and irritability when LLMs were unavailable, slightly ahead of the senior and mid-level cohorts on these specific items. Senior professionals reported the highest frequencies on items related to stress-driven use and difficulty stopping once started, standing out on these particular behaviors despite reporting comparatively lower frequencies elsewhere. Overall, these behaviors were reported across all experience levels, with each cohort showing a distinct concentration of behaviors rather than one group consistently reporting higher or lower frequencies across the full set of items.

Descriptions of how work would proceed without LLMs provide additional context for these findings. Emotional reactions were uncommon, with only a few respondents referring to the absence of LLMs as tedious," annoying," or draining" (e.g., P027, P086, and P089). In contrast, 35 out of 119 respondents indicated that the absence of these tools would have little or no effect on their work. For example, P005 stated, *Not much. I would switch to Google and Stack Overflow."* Rather than describing an inability to continue development activities, these responses emphasized the use of alternative resources, with changes limited to the tools employed rather than the overall workflow. These experiences suggest that addiction-related behaviors were not strongly reflected in everyday software engineering practice. Although the structured questions captured behaviors commonly discussed in the technology addiction literature, the professionals' comments rarely indicated loss of control, compulsive use, or substantial disruption to work.

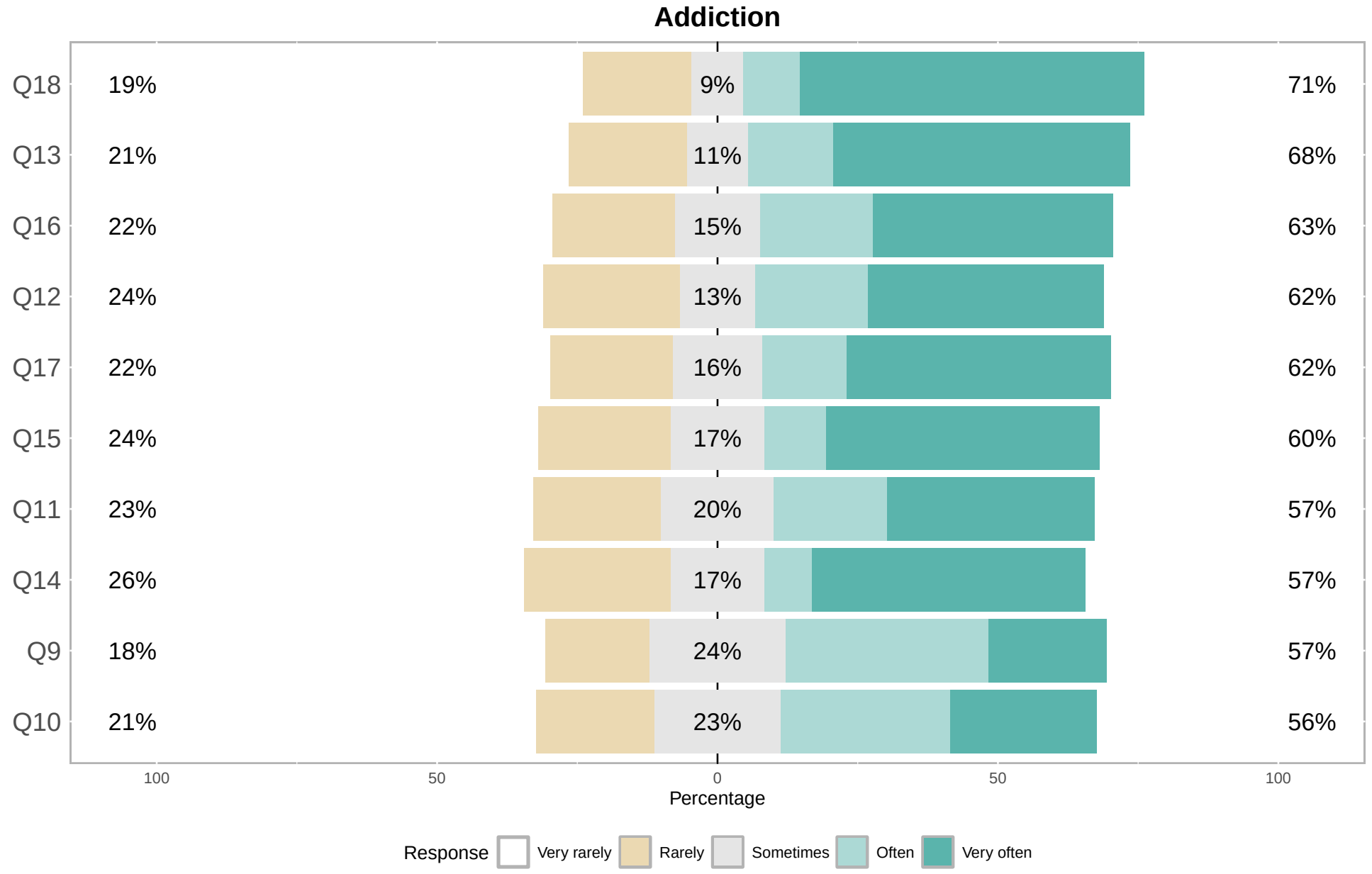

**Figure 6:** Responses to Likert-scale questions – addiction.

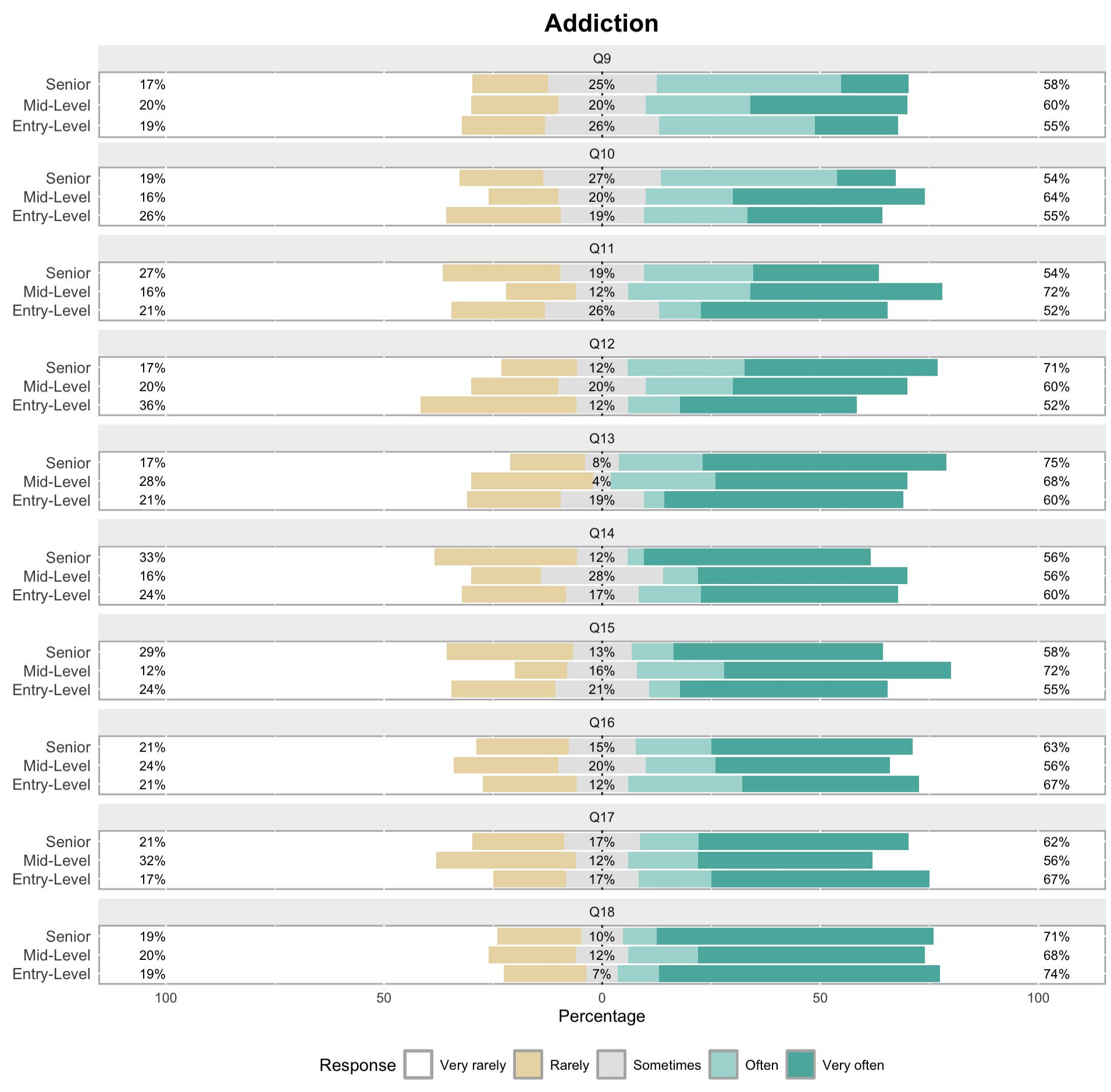

**Figure 7:** Responses to Likert-scale questions by experience level – addiction.

### 4.3. RQ: *How do software engineers experience the use of LLMs during professional software development, and what behavioral patterns associated with dependence, overreliance, and addiction-related behaviors emerge from their experiences?*

Across the findings, LLM usage was present in all experiences shared by the software engineering practitioners, but it is how they experience the absence of these tools during development work that separates their behaviors. Dependence emerged most clearly in reduced efficiency, with participants describing slower task completion and greater reliance on manual search when LLMs were unavailable. Overreliance, by contrast, was reflected in LLMs displacing peers and documentation as the primary source of technical support, with several participants describing these tools as their default approach to solving development problems and treating colleagues and documentation as fallback options rather than routine parts of their workflow. Addiction-related behaviors were the least supported by participants' accounts: most described only minimal disruption to their work, substituting alternative resources such as search engines or community forums rather than reporting compulsive use or an inability to continue engineering tasks. Software engineers' relationship with LLMs is therefore best characterized as habitual and preferential rather than compulsive, with dependence and overreliance as the dominant patterns and addiction-related behaviors present only marginally.

## 5. Discussion

This section discusses the findings in three parts. First, we compare our results with prior literature on LLM use and related behavioral patterns to situate the observed trends within existing knowledge. Next, we discuss the implications of these findings for software engineering practice, particularly in relation to how LLMs are integrated into daily work and how they may influence decision-making, collaboration, and skill development. Finally, we present lessons learned grounded in our results, focusing on practical considerations for the use of LLMs in professional environments.

### 5.1. Comparing Results with the Literature

Software engineering research has documented the use of LLMs across the development lifecycle, from requirements engineering to implementation, testing, debugging, documentation, and maintenance (Hou et al., 2024; Fan et al., 2023; Ozkaya, 2023; Gao et al., 2025; Haque, 2025; Zheng et al., 2025), showing that engineers perceive these tools as useful for productivity while also raising concerns about code quality, human oversight, and trustworthiness (Coutinho et al., 2024; Rasnayaka et al., 2024; Lo, 2023; Sallou et al., 2024). Our findings support this research: participants likewise described LLMs as integral to their daily work and reported substantial value in using them across engineering activities. In addition, our study extends this literature by showing that these benefits are accompanied by dependence, overreliance, and addiction-related behaviors, which this line of research has largely left unaddressed in software engineering. These behaviors have instead been studied primarily in the behavioral literature, in particular, considering students and general AI users (Zhai et al., 2024; Zhou and Zhang, 2024; Yankouskaya et al., 2024; Ciudad-Fernández et al., 2025), leaving practicing software engineers, a population whose profession is defined by the continual adoption of new tools and practices, largely unaddressed. In software engineering, adapting to new technologies is generally treated as an expected part of professional competence, and the behavioral cost of that expectation has received little attention. Our study investigates dependence, overreliance, and addiction-related behaviors in this population, connecting these two bodies of work and showing that integrating LLMs into daily engineering practice has behavioral implications beyond technical adoption. Therefore, the novelty of this study lies not in documenting that these behaviors occur, but in showing that they are not interchangeable: dependence, overreliance, and addiction-related behaviors reflect distinct ways software engineers integrate LLMs into their professional work, each with different implications for practice.

#### 5.1.1. *Dependence as Routine Integration of LLMs*

Our findings demonstrate that software engineers associated the absence of LLMs with increased effort and longer task completion times rather than an inability to perform software engineering activities. This observation is consistent with previous studies showing that LLMs reduce the effort required for coding, debugging, testing, and documentation while contributing to developers' productivity across a range of software engineering activities (Hou et al., 2024; Fan et al., 2023; Ozkaya, 2023; Haque, 2025; Zheng et al., 2025; Coutinho et al., 2024; Rasnayaka et al., 2024). Our findings extend this work by suggesting that these productivity gains are accompanied by a dependence pattern in which LLMs become incorporated into routine professional activities because they facilitate everyday software development.

This interpretation is consistent with the literature on functional dependence, which characterizes dependence as the routine incorporation of a technology into everyday activities because of its perceived usefulness rather than as evidence of pathological behavior (Fan et al., 2017). Within this perspective, dependence is viewed as a consequence of repeatedly relying on LLMs to accomplish work-related goals more efficiently. Likewise, the behavioral literature distinguishes this functional form of dependence from pathological dependence, emphasizing that repeated engagement alone is insufficient to indicate impaired control or problematic technology use (Chamberlain et al., 2016; Ciudad-Fernández et al., 2025).

#### 5.1.2. *Overreliance and Appropriate Reliance*

Overreliance among software engineers is expressed less through unquestioned acceptance of LLM-generated outputs and more through how these tools are incorporated into decision-making. Most respondents rejected deploying LLM-generated code without review, indicating that independent judgment and verification remain part of their practice. At the same time, many reported consulting LLMs before documentation or colleagues when seeking information or solving technical problems. This pattern aligns with the behavioral literature, which characterizes overreliance not by the amount of AI use but by trust calibration: users adjust their trust according to a technology's capabilities and limitations while maintaining independent judgment (Passi and Vorvoreanu, 2022; Kim et al., 2025; Klingbeil et al., 2024). The software engineering literature has similarly emphasized human oversight, critical evaluation, and verification when using LLMs because these systems may produce convincing but incorrect or incomplete outputs (Fan et al., 2023; Ozkaya, 2023; Gao et al., 2025; Sallou et al., 2024; Lo, 2023). However, these recommendations have generally been discussed from the perspective of software quality and reliability rather than through the concept of trust calibration. Our findings suggest that these perspectives are complementary. Verifying generated artifacts and consulting LLMs before alternative sources reflect different aspects of reliance, and both are important for understanding how these systems influence professional software engineering practice.

Appropriate reliance in software engineering therefore extends beyond verifying generated code before deployment and also involves determining when LLMs should support professional judgment and when other sources, including documentation, prior experience, or colleagues, should guide decision-making. Software engineers in our sample generally maintained appropriate reliance when evaluating generated outputs, but their information-seeking behavior was increasingly centered on LLMs. This suggests that appropriate reliance depends not only on validating AI-generated artifacts, but also on exercising professional judgment about when and how LLMs should be incorporated into everyday development activities.

#### 5.1.3. *Addiction and Practical LLM Use*

Our findings provide limited evidence that LLM use among software engineers is associated with addiction-related behaviors. Although many respondents reported behaviors related to time spent using LLMs, difficulty reducing use, or emotional reactions when access was unavailable, their descriptions of everyday work rarely reflected loss of control or substantial disruption to professional activities. Instead, participants primarily described the absence of LLMs as increasing the effort and time required to complete development tasks. This scenario suggests that, within professional software engineering, intensive LLM use is primarily associated with practical and goal-oriented work rather than compulsive engagement.

This interpretation is consistent with the behavioral literature, which characterizes addiction behavior as involving impaired control, persistence despite negative consequences, and meaningful disruption to everyday functioning rather than simply repeated or intensive technology use (Chamberlain et al., 2016; Marchica et al., 2022). Recent work has also cautioned against interpreting intensive interaction with conversational AI as evidence of addiction without demonstrating functional impairment or loss of control (Ciudad-Fernández et al., 2025; Yankouskaya et al., 2024). Our findings support this perspective within software engineering. While some responses to the structured survey questions were consistent with behaviors discussed in the addiction literature, participants' descriptions of their professional practice rarely reflected compulsive use or an inability to regulate their interaction with LLMs.

### 5.2. Implications

Our findings have implications for both research and practice by showing that the integration of LLMs into professional software engineering extends beyond technical performance and involves behavioral patterns that influence how developers interact with these tools in their everyday work.

#### 5.2.1. *Implications for Research*

Our study broadens current research on LLMs in software engineering in three ways. First, it provides a global perspective based on the experiences of software engineers from multiple countries, extending current discussions beyond productivity, software quality, and technical performance to behavioral aspects of LLM use in professional practice. While these behavioral perspectives have received attention in psychology and human-computer interaction, they have received limited attention in software engineering. Second, our findings shed light on questions that remain underexplored in software engineering. In particular, they suggest that understanding LLM adoption requires investigating not only verification and human oversight, but also trust calibration, that is, how software engineers determine when to rely on LLMs and when to seek alternative sources of information or exercise independent judgment. This perspective complements existing work on trustworthy AI by introducing behavioral mechanisms that may explain how developers interact with these systems in practice. Finally, our findings open opportunities for investigating the effects of LLMs beyond technical outcomes. Future research may investigate how long-term interaction with LLMs influences professional learning, decision-making, collaboration, software engineering expertise, well-being, and the development of work practices. Combining empirical software engineering with behavioral perspectives may contribute to a more comprehensive understanding of how LLMs influence software engineering over time.

#### 5.2.2. *Implications for Practice*

Our findings can support software companies in navigating the ongoing adaptations introduced by LLMs by providing concrete evidence on how these tools influence developers' work beyond technical performance. Our contribution lies specifically at the intersection of need and overuse, showing that software engineers' reliance on LLMs is driven by functional necessity rather than by patterns of compulsive use. Based on these findings, we derive the following lessons learned for the software industry:

- **Lesson 1: Recognize functional dependence as a natural consequence of productive LLM use.** Organizations should recognize that developers will naturally incorporate LLMs into routine software engineering activities because these tools reduce effort and improve productivity. Rather than discouraging this use, organizations should ensure that developers maintain the knowledge and skills required to complete tasks independently when necessary.
- **Lesson 2: Promote trust calibration and preserve professional judgment.** Verifying LLM-generated artifacts remains essential, but organizations should require developers to critically evaluate when LLMs are the appropriate source of support and when documentation, prior experience, or collaboration with colleagues should guide decision-making. When LLMs displace peers and documentation as the first point of reference, teams should implement strategies to preserve practices that maintain independent evaluation when designing solutions, reviewing generated artifacts, and making technical decisions.
- **Lesson 3: Distinguish intensive professional use from problematic technology use.** Frequent interaction with LLMs should not be interpreted as evidence of addiction without indications of impaired control or disruption to professional work. Organizations should instead monitor situations where LLM use begins to interfere with developers' autonomy, collaboration, or well-being.
- **Lesson 4: Tailor organizational support to developers' experience levels.** Organizations should avoid adopting a one-size-fits-all approach to LLM adoption. Developers at different career stages may require different forms of guidance, training, and support to promote appropriate reliance, preserve professional judgment, and encourage effective use of LLMs throughout software development.
- **Lesson 5: Incorporate behavioral considerations into organizational AI strategies.** Policies, training, and governance for LLM adoption should address not only technical issues such as software quality and security, but also how developers rely on LLMs, seek information, and integrate these tools into their everyday work.

# 6. Conclusion

In this study, we investigated how software engineers experience the use of LLMs during professional software development and identified behavioral patterns associated with dependence, overreliance, and addiction-related behaviors. Based on survey responses from 119 practitioners, our findings indicate that functional dependence is the predominant behavioral pattern, reflecting the routine incorporation of LLMs into everyday software engineering activities because of the productivity and efficiency these tools provide. Overreliance is primarily expressed through changes in how developers seek information and support, with LLMs often becoming the preferred starting point while professional judgment and verification remain present. In contrast, addiction-related behaviors were not strongly reflected in participants' professional practice, suggesting that intensive LLM use among software engineers is better understood as practical and goal-oriented rather than compulsive.

Our research contributes to software engineering research by extending current discussions on LLM adoption beyond technical capabilities and productivity. While previous work has primarily investigated the benefits, limitations, and technical implications of LLMs, our findings introduce a human perspective that connects software engineering with behavioral research. In particular, our results suggest that understanding how software engineers rely on LLMs requires considering not only verification and human oversight, but also functional dependence and trust calibration as complementary dimensions of professional AI adoption.

Our findings should be interpreted within the scope of this exploratory study. The objective was not to represent the entire software engineering population, but to identify behavioral patterns that may accompany the growing integration of LLMs into professional practice. Building on this study, our immediate next step is to conduct semi-structured interviews with software engineers to obtain a deeper understanding of how dependence, overreliance, and addiction-related behaviors are experienced in professional practice and how these behaviors develop over time. This investigation will provide a foundation for future studies across different organizational settings, software engineering roles, and development contexts, contributing to a more comprehensive understanding of how organizations can support effective LLM adoption while preserving professional judgment, software engineering expertise, and developers' well-being.

# CRediT authorship contribution statement

**Ronnie de Souza Santos:** Conceptualization, Methodology, Investigation, Formal analysis, Writing – original draft, Writing – review & editing. **Italo Santos:** Methodology, Investigation, Formal analysis, Writing – original draft, Writing – review & editing. **Matheus de Morais Leça:** Investigation, Writing – review & editing. **Cleyton Magalhães:** Writing – review & editing. **Mairieli Wessel:** Writing – review & editing.

## Declaration of generative AI and AI-assisted technologies in the manuscript preparation process

During the preparation of this work, the authors used generative artificial intelligence tools only for editorial support, language revision, and clarity improvement. After using these tools, the authors reviewed and edited the content as needed and take full responsibility for the content of the article.

## Declaration of competing interest

The authors declare that they have no known competing financial interests or personal relationships that could have appeared to influence the work reported in this paper.

## Ethics statement

This study received approval from the University of Calgary Research Ethics Board. Participation in the survey was voluntary, informed consent was obtained from all participants, and all responses were anonymized before analysis.

## Data availability

The data analyzed in this research are available at https://figshare.com/s/578016fd61528a0e1226